\RequirePackage{ifpdf}
\ifpdf % We are running pdfTeX in pdf mode
\documentclass[pdftex]{sigma}
\else
\documentclass{sigma}
\fi

\begin{document}

\def\a{\alpha}
\def\d{\delta}
\def\e{\epsilon}
\def\o{\omega}
\def\r{\rho}
\def\th{\theta}
\def\L{\Lambda}
\def\l{\lambda}
\def\i{\iota}
\def\t{\tau}
\def\s{\sigma}
\def\G{\Gamma}
\def\ca{{\cal A}}
\def\cb{{\cal B}}
\def\cf{{\cal F}}
\def\dim{\textrm{dim}}
\newcommand{\be}{\begin{equation}}
\newcommand{\ee}{\end{equation}}
\newcommand{\bea}{\begin{eqnarray}}
\newcommand{\eea}{\end{eqnarray}}

\renewcommand{\PaperNumber}{086}

\FirstPageHeading

\renewcommand{\thefootnote}{$\star$}

\ShortArticleName{Quantum Gravity as a Broken Symmetry Phase of a
BF Theory}

\ArticleName{Quantum Gravity as a Broken Symmetry Phase\\ of a BF
Theory\footnote{This paper is a contribution to the Proceedings of
the O'Raifeartaigh Symposium on Non-Perturbative and Symmetry
Methods in Field Theory
 (June 22--24, 2006, Budapest, Hungary).
The full collection is available at
\href{http://www.emis.de/journals/SIGMA/LOR2006.html}{http://www.emis.de/journals/SIGMA/LOR2006.html}}}

\Author{Aleksandar MIKOVI\'C~$^{\dag\ddag}$}
\AuthorNameForHeading{A. Mikovi\'c}

\Address{$^\dag$~Department of Mathematics, Lusofona University,\\
$\phantom{^\dag}$~Av. Do Campo Grande 376, 1749-024 Lisbon,
Portugal}
\EmailD{\href{mailto:amikovic@ulusofona.pt}{amikovic@ulusofona.pt}}

\Address{$^\ddag$~Mathematical Physics Group, University of Lisbon,\\
$\phantom{^\dag}$~Av. Prof. Gama Pinto 2, 1649-003 Lisbon,
Portugal}

\ArticleDates{Received October 02, 2006, in f\/inal form November
21, 2006; Published online December 07, 2006}

\Abstract{We explain how General Relativity with a cosmological
constant arises as a~broken symmetry phase of a BF theory. In
particular we show how to treat de Sitter and anti-de Sitter cases
simultaneously. This is then used to formulate a quantisation of
General Rela\-tivity through a spin foam perturbation theory. We
then brief\/ly discuss how to calculate the ef\/fective action in
this quantization procedure.}


\Classification{83C45; 81R50}

\section{Introduction}

Quantization of General Relativity (GR) is still one of the
outstanding problems in theoretical physics. Although the string
theory has made a signif\/icant progress on this problem
\cite{str}, the string theory approach is essentially a
perturbation theory around the f\/lat spacetime, which then makes
it dif\/f\/icult to study the quantum cosmology problems. On the
other hand, the Loop Quantum Gravity (LQG) approach is
nonperturbative and background metric independent \cite{lqg};
however, obtaining the perturbative and semiclassical results is
dif\/f\/icult. This is related to the fact that LQG is not
manifestly covariant under four-dimensional dif\/feomorphisms.
This problem has been overcome by introducing the spin foam
formalism \cite{sf}, where the basic object of LQG, the spin
network, a colored graph which lives in the space, is generalised
to a spin foam (SF), a colored two-complex which lives in the
spacetime. The SF models are typically obtained from the
path-integral quantization of a BF theory \cite{sf}, and the
reason for this is that the Einstein--Hilbert action in the
Palatini formalism
\[
 S_{EH} = \int_M \e^{abcd} e_a \wedge e_b \wedge R_{cd}, %\label{ehpa}
\]
\looseness=-1
 where the $e_a$ denote the tetrads and
$R_{ab} = d\o_{ab} + \o_{ac}\wedge\o^c_a$ is the curvature
two-form for the spin connection $\o_{ab}$ on the four-manifold
$M$, can be represented as a constrained $SO(3,1)$ BF theory
\[
S_{EH} = \int_M B^{ab}\wedge F_{ab},
\] where $F=R$ and the two-form $B$ is
constrained by the relation \be B^{ab} =\e^{abcd} e_c \wedge e_d
.\label{bco} \ee

The BF theories are topological, and their path-integral
quantization is well understood \cite{sf}. Let the BF theory group
be a Lie group $G$ and let $\L$ label the f\/inite-dimensional
irreducible representations (irreps) of $G$, then the partition
function can be written as a sum over the irreps of the spin foam
amplitudes associated to the dual 2-complex $\G$ of a
triangulation of $M$ \be Z_{BF}=\sum_{\L_1,\dots
,\L_F}\sum_{\i_1,\dots ,\i_L} \prod_{f=1}^F \dim\,\L_{f}
\prod_{l=1}^L A_{\t}(\L_{f(l)} ) \prod_{v=1}^V A_{\s} (\L_{f(v)}
,\i_{l(v)}),\label{bfss} \ee where the $\L$'s label the faces $f$
of $\G$, the $\i$'s are the corresponding intertwiners which label
the edges $l$ of $\G$, $A_{\t}$ are the tetrahedron amplitudes and
$A_{\s}$ are the four-simplex amplitudes. The~$A_l$ is a function
of the dimensions of the four irreps that meet at an edge $l$,
while $A_v$ is given by an evaluation of the four-symplex spin
network where the ten edges carry the irreps $\L_f$ and the f\/ive
vertices carry the intertwiners $\i_l$ \cite{sf}. The inf\/inite
sums in (\ref{bfss}) are divergent and the typical regularization
is to replace the Lie group $G$ by the quantum group $G_q$ where
$q$ is a root of unity~\cite{cky}. In this case the set of $\L_f$
becomes f\/inite so that (\ref{bfss}) becomes f\/inite.

However, the quantization of constrained BF theories is not that
well understood, and in the GR case there is a proposal by Barrett
and Crane \cite{bce,bcl} to implement the constraint (\ref{bco})
in the SF formalism by restricting the irreducible representations
of $SO(3,1)$ that color the faces of the spin-foam complex as \be
\e^{abcd}J^{(\L)}_{ab}J^{(\L)}_{cd} = 0,\label{srep} \ee where
$J^{(\L)}$ are the Lorentz group generators in the representation
$\L$. Furthermore, one can argue that the admissible irreps have
to be unitary, so that (\ref{srep}) selects the so called simple
unitary irreps of the Lorentz group $\L=(j,0)$ or $\L=(0,\rho)$
where $2j\in {\mathbb Z}_+$ and $\rho\in(0,\infty)$ \cite{bcl}.
Given that a triangulation of a four-dimensional pseudo-Riemannian
manifold can be always chosen such that all the triangles are
spacelike, one can work with only the $(0,\rho)$ irreps. In this
way one obtains a spin foam model where the GR path integral is
def\/ined as a multiple integral over the face irreps of the
corresponding spin foam amplitudes and (\ref{bfss}) becomes
\[
Z_{GR}=\int_0^{\infty}d\r_1 \cdots\int_0^\infty d\r_F
\prod_{f=1}^F \r_f \prod_{l=1}^L \th^{-1}(\r_{f(l)})\prod_{v=1}^V
A_\s^{\rm rel} (\r_{f(v)}),
\]
where $\th$ is a function of the four edge irreps and $A_\s^{\rm
rel}$ is the relativistic evaluation of a four-simplex spin
network.

One can show that for each non-degenerate triangulation of $M$ the
corresponding BC spin foam state sum is f\/inite, provided that
the $\th$ amplitudes are appropriately chosen \cite{cpr}. In this
way one obtains a f\/inite theory of quantum gravity; however,
there are problems with this theory. First, the choice of the
$\th$ amplitudes is not unique, since there are other choices
which also lead to a f\/inite state sum \cite{bacr}. Second,
coupling fermionic matter is dif\/f\/icult because fermion
f\/ields couple to individual tetrads, while in the BC model one
can couple the fermions only to a~specif\/ic quadratic combination
of the tetrads (the B f\/ield). And the last, and the most
dif\/f\/icult problem is to f\/ind the semiclassical limit of this
quantum gravity theory.

\section{GR as a symmetry breaking of a BF theory}

The f\/irst two problems of the BC model suggest that one should
look for a spin foam model of GR which arises from the
quantization of a BF theory which is not constrained and includes
the tetrads in the BF theory connection one-form. This also means
that GR will appear through a symmetry breaking mechanism, since
the larger symmetry of a bigger connection has to be broken to a
Lorentz group connection plus the tetrads. Interestingly, such a
mechanism was found by MacDowell and Mansouri \cite{mm} in the
context of $OSp\,(n|4)$ supergravity theories. The bosonic spatial
symmetry group was $Sp(4)$, which is the covering group of the
anti-de Sitter group $SO(3,2)$. Their action was quadratic in the
f\/ield strengths and gave EH action with a~negative cosmological
constant. A BF theory formulation for a positive cosmological
constant EH action was found by Smolin and Starodubtsev \cite{ss},
and  corresponds to the de Sitter group case, i.e.\ $SO(4,1)$.

Extending the Smolin--Starodubtsev result to the anti-de Sitter
case is easy, since one can treat the both cases simultaneously
in the following way. The Lie algebras of both groups can be
represented as
\[
[J^{ab},J^{cd}]=\eta^{[a[c}J^{d]b]},\qquad
[J^{ab},P^c]=-\eta^{c[a}P^{b]} ,\qquad [P^a,P^b]=\pm J^{ab},
\]
where $\pm$ corresponds to anti-de Sitter and de Sitter cases
respectively. The corresponding connection can be written as
\[
 {\cal A} = \o_{ab} J^{ab} + \l e_a P^a ,
\]
where $\o$ is the spin-connection, $e_a$ are the tetrads and $\l$
is a dimensionful parameter. The curvature 2-form ${\cal F} =
d{\cal A}+{\cal A}\wedge{\cal A}$ is then given by
\[
{\cal F} = T_a P^a + (R_{ab} \pm \l^2 e_a \wedge e_b )J^{ab},
\]
where $T_a = de_a + \o_a^b \wedge e_b$ is the torsion. Let us
introduce the Lie algebra valued 2-form
\[
 {\cal B}=b_a P^a + B_{ab}J^{ab} ,
\]
then consider the following action \be S=\int_M \left(Tr\,({\cal
B} \wedge {\cal F} ) -\frac{\a}{2}
 \e^{abcd}B_{ab}\wedge B_{cd}\right).\label{dbf}
 \ee
This is the BF theory action which is perturbed by a symmetry
breaking term, since the quadratic in B term is invariant only
under the Lorentz subgroup. This action can be written as
\[
S=\int_M \left( b^a\wedge T_a + B^{ab} \wedge (R_{ab}\pm \l^2 e_a
\wedge e_b) -\frac{\a}{2} \e^{abcd}B_{ab}\wedge B_{cd}\right),
\]
so that the $b$ and $B$ equations of motion imply vanishing of the
torsion and
\[
B^{ab}=\frac{1}{\a}\e^{abcd}(R_{cd}\pm \l^2 e_c \wedge e_d).
\]
The remaining equations correspond to the action
\[
S^* = \frac{1}{2\a}\int_M \e^{abcd}(R_{ab}\pm \l^2 e_a \wedge
e_b)\wedge (R_{cd}\pm \l^2 e_c \wedge e_d).
\]

The plus sign gives the MacDowell--Mansouri action, while the
minus sign corresponds to the theory studied in~\cite{ss}. Either
way, it is easy to see that $S^*$ can be written as
\[
S^* =\pm\frac{1}{G_N}\int_M \e^{abcd}\,e_a \wedge e_b \wedge
(R_{cd} \pm \L e_c \wedge\e_d ) + S_{\rm top} ,
\]
where $G_N =\a \l^{-2}$ is the Newton constant and $\L =
\frac{\l^2}{2}$ is the cosmological constant and $S_{\rm top}$ is
proportional to the Euler class of $M$, which does not af\/fect
the equations of motion. Hence one obtains the Einstein equations
for positive/negative cosmological constant. Since $\a=2G_N \L$,
this gives that $\a$ is an extremally small number, which then
justif\/ies the view that GR with a~small cosmological constant is
a~small deformation of a BF theory.

\section{Spin foam perturbation theory}

The action (\ref{dbf}) is well-suited for a perturbative
quantization. The standard approach is to calculate perturbatively
the generating functional
\[
 Z[j,J] =\int {\cal D}\ca\, {\cal D}\cb
\exp\left( i\int_M  {\rm Tr}\left(\cb\wedge \cf  +j\wedge \ca+ J\wedge \cb\right)+U(\ca,\cb)\right),%\label{papb}
\]
where $J$ and $j$ are the sources for the $\ca$ and the $\cb$
f\/ields and $U(\ca,\cb)$ is the perturbative interaction. This
can be done via the formula
\[
Z[j,J]=\exp\left( i\int_M U\left(\frac{1}{i}\frac{\d}{\d
j},\frac{1}{i}\frac{\d}{\d J}\right)\right)Z_0 [j,J],
%\label{pertz}
\]
where
\[
Z_0 [j,J] =\int {\cal D}\ca {\cal D}\cb\exp\left(i\int_M \,{\rm
Tr}\left(\cb\wedge \cf
+ j\wedge \ca + J\wedge \cb\right)\right).%\label{genf}
\]

This generating functional can be calculated by using the spin
foam technology \cite{sard}, and the result can be written as a
state sum \be Z_0 [j,J] = \sum_{\L_f ,\l_l , \i_l} \prod_f \dim
\L_f \Big{\langle} \prod_l \mu (\l_l ,j_l ) \prod_v A_\s
(\L_{f(v)}, \l_{l(v)},\i_{l(v)}, J_{f(v)})
 \Big{\rangle},\label{sfgf}
 \ee
where
\[
\mu (\l_l ,j_l ) = \int_G dg_l \left(D^{(\l_l)}(g_l)\right)^*
e^{i\,{\rm Tr}\,(A_l j_l)} ,
\]
is the insertion at an edge $l$ of $\G$ while $A_\s (\l,J)$ is the
modif\/ied 4-symplex amplitude, with the $\l$-edges attachments at
its vertices and $D^{(\L_f)}(e^{J_f})$ insertions at its edges.
When the sources vanish, the state sum (\ref{sfgf}) reduces to the
state sum (\ref{bfss}). Notice that the state sum (\ref{sfgf}) is
not of the same type as (\ref{bfss}), because one has to label
both the edges and the faces of $\G$ with the irreps of $G$. The
expression (\ref{sfgf}) is an inf\/inite sum, and has to be
regularized. When $G=SO(5)$, i.e.\ the Euclidian gravity case, the
regularization consists of replacing the category of irreps of $G$
by the category of irreps of the quantum group $G_q$ where $q$ is
a root of unity. When $G=SO(3,2)$ or $G=SO(4,1)$, one has to use
the category of unitary irreps, which are inf\/inite-dimensional
and typically have discrete and continuos series. This means that
one will obtain both the state sums and the state integrals (like
in the BC model case). Convergence properties of these models have
not been yet investigated, and the hope is that the corresponding
categories of quantum group irreps will yield convergent sums.
Alternatively, one could try to generalize the discrete gauge
f\/ixing procedure introduced by Freidel and Louapre for the
three-dimensional spin foam models \cite{fl,fl2}.

Given a regularized $Z_0 [j,J]$ one can calculate the generating
functional perturbatively as
\[
Z[j,J] = Z_0 [j,J]+ \a Z_1 [j,J] + \a^2 Z_2 [j,J] + \cdots.
\]
The semiclassical properties of the theory where $Z_0$ is given by
(\ref{sfgf}) can be explored by analyzing the ef\/fective action.
The ef\/fective action can be calculated via the Legendre
transform of $Z[J,j]$
\[
\G(\bar\ca_l , \bar\cb_f ) = W(j_l ,J_f ) - \sum_l {\rm Tr}\,(j_l
\bar\ca_l ) - \sum_f {\rm Tr}\,(J_f \bar\cb_f ),
\]
where
\[
\bar\ca_l ={\partial W\over \partial j_l},\qquad  \bar\cb_f
={\partial W\over \partial J_f }, \qquad iW(j_l ,J_f ) = \log Z
(j_l ,J_f ).
\]
This can be done perturbatively in $\a$, and if we denote
$(\bar\ca_l ,\bar\cb_f )$ as $X_I$, then
\[
\G(X) = \sum_{m\ge 0}\a^m \sum_{n\ge 0}
\frac{1}{n!}\sum_{I_1,\dots, I_n} C_{mn}(I_1 \cdots I_n )
X_{I_1}\cdots X_{I_n}=\sum_{m\ge 0}\a^m \G_m (X) .
\]

In this way one can explore the semiclassical limit of the theory,
and the crucial test for the physical relevance of the theory is
whether or not $\G_0 +\a \G_1$ gives the discretized classical
action~(\ref{dbf})
\[
 S =  \sum_f {\rm Tr}\,({\cal B}_f {\cal F}_f)- \frac{\a}{2}\sum_{f,f'}C(f,f')\e^{abcd}B^{ab}_f B^{cd}_{f'}.
 \]

 \vspace{-3mm}

\section{Conclusions}

The spin foam perturbation theory is a promising approach for
def\/ining a viable quantisation of GR. However, one must resolve
f\/irst the technical questions related with the regularization of
the SF generating functional. This is something which may be
complicated, but it can be done. Then one can calculate the
ef\/fective action by the method outlined here and verify the
classical limit. If this limit is GR, one can then proceed to
calculate the higher order in $\a$ corrections. The matter can be
coupled by using again the same method of the generating
functional \cite{sard}.

On the mathematical side, one can explore the state sum
(\ref{sfgf}) for the case of compact groups and study what happens
with the topological invariance. In the non-compact group case,
one will need f\/irst to study the unitary irreps of quantum de
Sitter and anti-de Sitter groups. One can also study $Z_0 (j,J)$
for three-dimensional spacetimes, in which case the relevant
groups are $SO(4)$, $SO(3,1)$, $SO(2,2)$ and their covering
groups. In the case of two-dimensional spacetimes the relevant
groups are $SO(3)$, $SO(2,1)$ and their covering groups.

\vspace{-2mm}

\subsection*{Acknowledgements}

This work has been supported by the FCT grant
POCTI/MAT/45306/2002.

\LastPageEnding


\begin{thebibliography}{99}
\footnotesize

\bibitem{str} Green M.B., Schwarz J.H., Witten E., Superstring theory, Vols.~1,~2, Cambridge University Press, 1987.

\bibitem{lqg} Rovelli C., Loop quantum gravity, {\it Living Rev. Relativ.}, 1998, V.1, 1998-1, 68 pages,
\href{http://arxiv.org/abs/gr-qc/9710008}{gr-qc/9710008}.

\bibitem{sf} Baez J.C.,
An introduction to spin foam models of $BF$ theory and quantum
gravity, in Geometry and Quantum Physics (1999, Schladming), {\it
Lecture Notes in Phys.}, Vol.~543, Berlin,  Springer, 2000,
25--93, \href{http://arxiv.org/abs/gr-qc/9905087}{gr-qc/9905087}.

\bibitem{cky} Crane L., Kauf\/fman L.H., Yetter D., State-sum invariants of 4-manifolds,
{\it  J. Knot Theory Ramifications}, 1997, V.6, 177--234,
\href{http://arxiv.org/abs/hep-th/9409167}{hep-th/9409167}.

\bibitem{bce} Barrett J.W., Crane L., Relativistic spin networks and quantum gravity,
{\it J. Math. Phys.}, 1998, V.39, 3296--3302,
\href{http://arxiv.org/abs/gr-qc/9709028}{gr-qc/9709028}.

\bibitem{bcl} Barrett J.W., Crane L., A Lorentzian signature model for quantum general relativity,
{\it Classical Quantum Gravity}, 2000, V.17, 3101--3118,
\href{http://arxiv.org/abs/gr-qc/9904025}{gr-qc/9904025}.

\bibitem{cpr} Crane L., Perez A., Rovelli C., Perturbative f\/initeness in spin-foam quantum gravity,
{\it Phys. Rev. Lett.}, 2001, V.87, 181301, 4 pages.

\bibitem{bacr} Baez J.C., Christensen J.D., Halford T.R., Tsang D.C., Spin foam models of Riemannian quantum gravity,
{\it Classical Quantum Gravity}, 2002, V.19, 4627--4648,
\href{http://arxiv.org/abs/gr-qc/0202017}{gr-qc/0202017}.

\bibitem{mm} MacDowell S.W., Mansouri F.,
 Unif\/ied geometric theory of gravity and supergravity, {\it Phys. Rev. Lett.}, 1977, V.38, 739--742.

\bibitem{ss} Smolin L., Starodubtsev A., General relativity with a topological phase:
an action principle,
\href{http://arxiv.org/abs/hep-th/0311163}{hep-th/0311163}.

\bibitem{sard} Mikovi\'c A., Quantum gravity as a deformed topological quantum f\/ield theory,
 {\it J. Phys. Conf. Ser.}, 2006, V.33, 266--270,  \href{http://arxiv.org/abs/gr-qc/0511077}{gr-qc/0511077}.

\bibitem{fl} Freidel L., Louapre D., Dif\/feomorphisms and spin foam models,
{\it Nuclear Phys. B}, 2003, V.662, 279--298,
\href{http://arxiv.org/abs/gr-qc/0212001}{gr-qc/0212001}.

\bibitem{fl2} Freidel L., Louapre D., Ponzano--Regge model revisited.
I.~Gauge f\/ixing, observables and interacting spinning particles,
{\it Classical Quantum Gravity}, 2004, V.21, 5685--5726,
\href{http://arxiv.org/abs/hep-th/0401076}{hep-th/0401076}.

\end{thebibliography}
\end{document}